# The unpaved road towards efficient selective breeding in insects for food and feed

**Running title: Towards selective breeding in insects**

L.S. Hansen[1,3,*] (ORCID: 0000-0002-4270-6365)

S.F. Laursen[2] (ORCID: 0000-0002-7879-8660)

S. Bahrndorff[2] (ORCID: 0000-0002-0838-4008)

J.G. Sørensen[3] (ORCID: 0000-0002-9149-3626)

G. Sahana[1] (ORCID: 0000-0001-7608-7577)

T.N Kristensen[2] (ORCID: 0000-0001-6204-8753)

H.M. Nielsen[1] (ORCID: 0000-0002-8001-5629)

[1] Center for Quantitative Genetics and Genomics, Aarhus University, C.F. Møllers Allé 3, 8000 Aarhus, Denmark

[2] Department of Chemistry and Bioscience, Aalborg University, Fredrik Bajers Vej 7H, 9220 Aalborg East, Denmark

[3] Department of Biology, Aarhus University, Ny Munkegade 114, 8000 Aarhus, Denmark

[*]Correspondence: lsh@qgg.au.dk

**Keywords:** Insect breeding; phenotyping; selection; genetic improvement; genetic parameters

## Abstract

Insect production for food and feed presents a promising supplement to ensure food safety and address the adverse impacts of agriculture on climate and environment in the future. However, optimisation is required for insect production to realise its full potential. This can be by targeted improvement of traits of interest through selective breeding, an approach which has so far been underexplored and underutilised in insect farming. Here we present a comprehensive review of the selective breeding framework in the context of insect production. We systematically evaluate adjustments of selective breeding techniques to the realm of insects and highlight the essential components integral to the breeding process. The discussion covers every step of a conventional breeding scheme, such as formulation of breeding objectives, phenotyping, estimation of genetic parameters and breeding values, selection of appropriate breeding strategies, and mitigation of issues associated with genetic diversity depletion and inbreeding. This review combines knowledge from diverse disciplines, bridging the gap between animal breeding, quantitative genetics, evolutionary biology, and entomology, offering an integrated view of the insect breeding research area and uniting knowledge which has previously remained scattered across diverse fields of expertise.

## Summary

Insect farming can boost food security and reduce agriculture's impact on the climate and environment, but it needs optimisation. Selective breeding, although not widely used in insect farming, holds potential for improvement. This review examines how to apply selective breeding to insects. It discusses key steps like setting breeding goals, measuring traits, estimating genetic values, choosing breeding strategies, and managing genetic diversity. The review integrates insights from animal breeding, genetics, evolutionary biology, and entomology, providing a comprehensive guide to improving insect production through selective breeding.

## Introduction

The rapid expansion of the human population poses a pressing challenge: ensuring an adequate food supply to meet the growing demand while simultaneously reducing negative impacts of production on the environment, climate and biodiversity (FAO 2009). One critical aspect of this challenge lies in the increasing demand for protein to provide food for a rapidly growing human population. Traditional high-input, resource-intensive farming systems confronts great challenges e.g. due to its substantial contribution to greenhouse gas emissions and environmental footprints (Searchinger et al. 2018, Olesen et al. 2021). Therefore, there is an imperative need for innovative approaches that can augment food production with less negative impacts on climate and the environment (Mannaa et al. 2024). One promising strategy to mitigate the impending food crisis is the commercial production of insects as food or feed (van Huis & Gasco 2023). This proposition arises from the fact that insect protein can be produced in a more environmentally sustainable manner when compared to conventional animal protein sources (Smetana et al. 2016, van Huis et al. 2020, Laganaro et al. 2021, Lange & Nakamura 2023), which aligns with the principles of the circular economy paradigm on optimising resource utilisation and enhancing energy efficiency (IPIFF 2019).

While ongoing efforts in insect production show promise, it is imperative to optimise production methods to fully harness the potential benefits of insect farming. The optimisation of insect production can be achieved through various means, including improvements in environmental conditions such as rearing substrate, population density, and abiotic factors like temperature and humidity (van Huis & Tomberlin 2017). Another means for production optimisation involves genetic improvement of targeted traits within insect populations through selective breeding (Eriksson & Picard 2021) (Figure 1). Selective breeding involves the deliberate selection of breeding animals based on phenotypic characteristics of defined traits or the predicted genetic values. This practice distinguishes itself from experimental evolution, where evolutionary changes occur as a result of environmental, social, or demographic conditions (Kawecki et al. 2012). It is worth noting that, in some instances, the term "experimental evolution" is employed in scientific literature to describe the process of artificial selection (Armitage & Siva-Jothy 2005). Since experimental evolution does not imply any targeted selection, we use artificial selection synonymously with selective breeding.

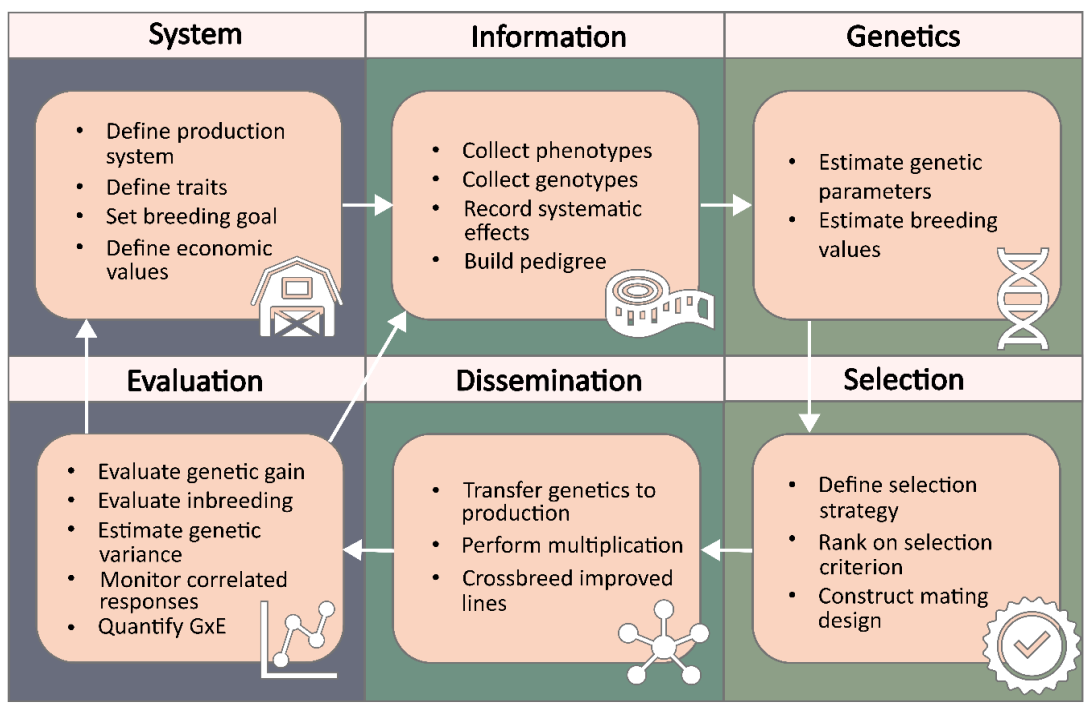

**Figure 1** Schematic overview of a generic breeding program. Prior to selection the production system and breeding objectives are defined. In each selection round information is collected, genetic evaluation and selection is performed, the genetic material is disseminated from the breeding population to the production unit, and the outcomes of selection are evaluated. This evaluation can result in an update of the breeding objective, or another selection round can be initiated.

Given the short generation interval and high fecundity of most insect species, it is reasonable to anticipate even more rapid genetic advancements within a comparable timeframe compared to what is observed in livestock species with longer generation intervals (Hill 2010). A few examples of selective breeding in insects exist. Facchini et al. (2022) reported results from an industrial genetic improvement program aimed at increasing larval body weight in black soldier fly (*Hermetia illucens*) larvae. After 16 generations of selection, they observed a remarkable 39% increase in larval weight. Similarly, Morales-Ramos et al. (2019) demonstrated an increase from 107 mg to 177 mg in individual pupal weight of yellow mealworm (*Tenebrio molitor*) after eight years of selection. These findings offer promise for the future of insect breeding. The selection strategy used in the studies rely on the direct observation on an individual's or group's phenotype. This has consequences for the accuracy of selection and does not allow accurate monitoring of inbreeding and underlying genetic correlations between traits, which can lead to undesired selection outcomes. For these reasons, it is desirable to explore more advanced breeding methodologies that employ selection based on the underlying genetic merit.

The establishment of complex breeding programs entails several critical steps (**Figure 1**). These include the identification of relevant traits to be selected for, the collection of phenotypic data on individuals or related group members, the estimation of genetic parameters, the maintenance of information on relatedness, and the creation of a breeding program that maximises multi-trait genetic progress while mitigating genetic drift and inbreeding. However, integrating insects into this selective breeding framework poses unique challenges. Therefore, it is essential to tailor breeding strategies to suit insect life history characteristics and evaluate each step of the breeding scheme within the context of commercial insect production.

This review aims to evaluate, discuss, and present the framework necessary for implementing selective breeding in insects intended for food and feed production. We meticulously assess each prerequisite in a breeding scheme and review the current advancements made in the context of insects. Furthermore, we analyse the opportunities and address the challenges posed by the most abundant commercial insect species within the context of selective breeding. Finally, we provide practical suggestions for approaching and surmounting these challenges. Our primary focus lies on large-scale industry settings and, consequently, on insect species presently utilised in such settings. We hypothesise that the genetic principles successfully applied in breeding schemes for livestock, aquaculture, and plant breeding can be adapted to insect populations, ultimately facilitating the optimisation of insect breeding to yield populations with desired trait characteristics. This study bridges the gap between traditional selective breeding practices and the emerging field of insect farming, offering valuable insights and methodologies that will aid the effectiveness of commercial production of insects for food and feed purposes.

## The basis of selective breeding

The evolution of complex traits is governed by a multifaceted interplay between genetics and the environment, which is studied using quantitative genetics and statistical approaches. The universality of genetic principles and inheritance mechanisms in diploid organisms facilitate the transferability of these methods across species and populations. The rate and direction of quantitative trait evolution can be influenced through selective breeding by influencing components of the Breeder's equation (Figure 2).

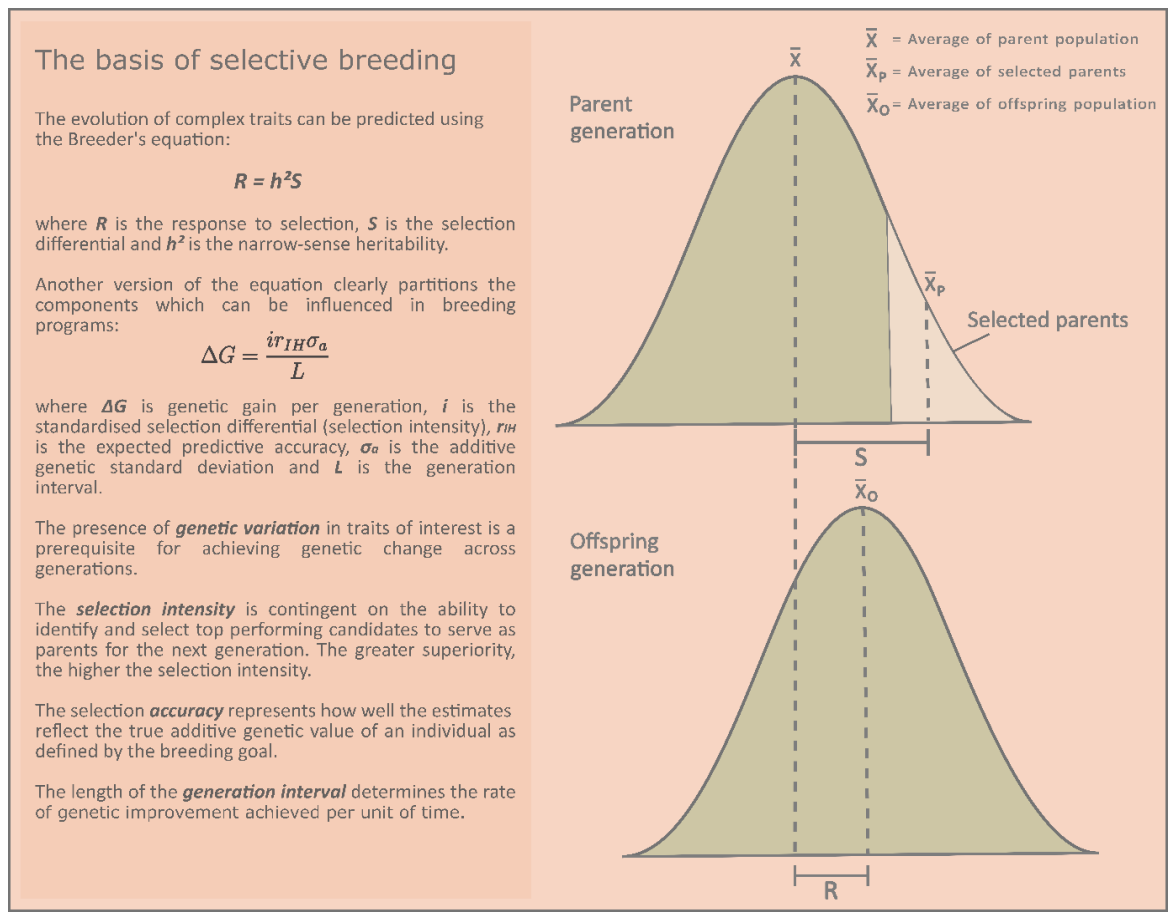

**Figure 2** The basis of selective breeding and a visualisation of the transgenerational result of selection.

While selection intensity and accuracy can be manipulated through design decisions in the breeding program, the initial presence of adequate additive genetic variance defines the potential for long-term genetic progress. As selection proceeds, genetic variance is reduced (Bulmer 1971) and selection becomes increasingly inefficient. Emphasis should be placed on establishing base populations with a broad genetic base, e.g. by crossing genetically distinct lines or sampling from wild populations of diverse geographical origins. Additionally, the phenotypic performance of several lines or populations should ideally be evaluated to identify the one(s) most fit for purpose (Adamaki-Sotiraki et al. 2022, Sandrock et al. 2022) and to establish a reference point for selection responses. When considering insect species of interest for food and feed production, these are often widely distributed in nature across vast geographical distances, where they have adapted to diverse environments (Rozkosný 1982, Marshall et al. 2015, Lessard et al. 2019, Delclos et al. 2021). Thus, their performance should be evaluated in the target production environment.

## Breeding goal and economic values

The desired direction of genetic improvement in a given population is expressed using a breeding goal. The breeding goal is specific for the given species and production system (e.g. a black soldier fly farm) and the production circumstances under which the insects are being produced (e.g. price scheme for protein and fat). It consists of several traits, where each trait is weighed according to its economic importance (economic value, Hazel 1943). Desired characteristics with economic relevance need to be identified in the early stages of the breeding program design, which require basic knowledge of the biology of the species in question. An economic value of a given trait is defined as a change in profit of the production system by a unit change in the trait.

For livestock species, bio-economic models have been broadly used to derive economic values (see review by Nielsen et al. 2014). A bio-economic model describes the production system by simulating biological factors such as survival, reproduction, and performance level, the cost of input such as feed and labour, and the output product of the system. In the case of insects, the production system could be a factory producing protein from insects, a breeding company aiming at selling genetically improved material to insect producers, or a fully integrated industry with both selective breeding and production of insect protein and oil. Derivation of economic values using a bio-economic model requires biological and economic data. Zaalberg et al. (2024) presented a bio-economic model to estimate economic values for traits in black soldier fly for a factory producing 1,000 kg of larvae harvested on day 15 of development. In that study, composition traits (dry mass and protein content) and growth traits (larval mass, development time and growth rate) had the highest economic values. In cases where economic data is unavailable, a restricted or desired gain index may be used (Groen et al. 1997) to derive economic values based on a desired genetic change in a given trait. Restricted and desired gain indices are based on a selection index, wherefore knowledge about genetic and phenotypic parameters are needed. Since in this case, economic values are derived using usually an arbitrarily chosen change in a given trait, there is no guarantee that economic values from the desired gain index are the ones yielding highest profit for the system (Gibson & Kennedy 1990).

## Insect biology in a selective breeding context

Insects constitute one of the most remarkably diverse animal groups, comprising an estimated 5.5 million species (Stork 2018). A small subset of these species has traditionally served as a source of food or feed on a global scale (e.g., Orkusz 2021). Within the European Union (EU), only a restricted subset of eight insect species has received official approval for use as feed, and four as novel food (Delgado Calvo-Flores et al. 2022). Notably, these insects exhibit markedly distinct characteristics (Figure 3) compared with conventional domesticated animals such as pigs, chickens, and cattle. These differences are of fundamental significance with respect to selective breeding efforts. Applying selective breeding requires manipulation and management of the entire rearing and breeding cycle. This includes manipulation of e.g. the rearing environment, mating, reproduction including reproductive behaviour, and longevity, all of which are intricately entwined with the insect biology (Morales-Ramos et al. 2024). In the following sections we discuss how the challenges encountered in this endeavour may differ across candidate insect species.

***Domestication***

Although the production of insects for food and feed has a comparatively short history in contrast to the longstanding traditions of livestock and crop domestication, it is worth noting that some insects have been farmed for centuries, thereby undergoing domestication. Prominent examples of mass-reared and domesticated insects include the honeybee (*Apis mellifera*) (Oldroyd 2012), silkworm (*Bombyx mori)* (Sun et al. 2012), and the lac insects (*Kerria* spp.) (Bashir et al. 2022). Examples of recent domestication histories include species used for waste management, animal feed production, and food production, such as the black soldier fly (Rhode et al. 2020), yellow mealworm (Eriksson et al. 2020), and house cricket (*Acheta domesticus)* (Lecocq 2018). Although some insects are better suited for rearing in captivity and selective breeding due to their ease of rearing and high productivity, the references above show evidence that it is feasible to achieve successful captive rearing of species originating from diverse taxonomic backgrounds. As the insect farming industry continues to mature, lessons from domestication history can guide efforts to expand the range of insect species reared for various purposes.

***Thermal environment***

The most fundamental distinction between insects and other terrestrial livestock lies in their thermoregulation strategy. Insects are ectotherms, thereby heavily reliant on their surrounding thermal conditions to influence a myriad of traits (Hoffmann et al. 2003, Sørensen et al. 2003, Tomberlin et al. 2009, Régnière et al. 2012, Shumo et al. 2019, Opare et al. 2022). Despite the ability to control temperature conditions in laboratory or production settings, individuals within populations will experience spatial heterogeneity in those conditions, which may complicate the separation of heritable genetic effects from non-heritable environmental effects (Pincebourde et al. 2007, Kearney et al. 2009, Li et al. 2023). Specific elements of the breeding program, such as rearing families in separate downscaled environments to ease tracking, can also have profound effects on the thermal environment within the rearing containers (Yakti et al. 2022), possibly influencing traits related to development, growth, behaviour, and fitness. Such effects should be investigated before initiating a breeding program.

***Life cycle***

All insects share a common characteristic in their development, characterised by discrete stages during which they periodically shed their exoskeleton. Insects can be broadly categorised into two main groups: holometabolous and hemimetabolous species (Figure 3). Holometabolous insects undergo one or more larval stages followed by a pupal stage, during which the adult insect develops through complete metamorphosis. This is the case for farmed species such as the black soldier fly, yellow mealworm, silkworm, and house fly (*Musca domestica*). In contrast, hemimetabolous insects (e.g. house crickets, locusts (order: Orthoptera), and cockroaches (order: Blattodea)) progress through a series of nymphal stages that increasingly resemble the adult form, exhibiting incomplete metamorphosis. Both types of development encompass substantial differences in abiotic requirements, physiology, and tolerances across various life stages (Tomberlin & Sheppard 2002). The generation interval in insects is highly contingent on factors such as temperature and other abiotic conditions (Mirth et al. 2021). The generally swift pace of development introduces practical implications: eggs will hatch, insects will progress to subsequent life stages, surpass their reproductive windows, or even die before breeding decisions can be enacted. Time-intensive tasks, such as phenotyping and genetic evaluations, must, therefore, be tailored to align with the life cycle of the species under consideration. High-throughput methodologies, workflows and technological advances are needed to effectively accommodate these temporal constraints.

***Mating system and reproduction***

In animal breeding, conventional practices often involve a straightforward procedure of mating, where males and females engage in sexual reproduction under highly controlled conditions or reproduce by artificial insemination (one exception being several fish species such as sea bass and sea bream, which rely on mass spawning (Superio et al. 2021). Application of artificial insemination is not practical in insects used for commercial food and feed production, with the notable exception of honeybee queen production (Khan et al. 2022). In insects, four potential mating systems exist that significantly affect the ability to manage mating: monogamy and three types of polygamy (polygyny, polyandry, and polygynandry) (e.g., Hoffmann et al. 2021). While there are instances of strict monogamous insects (Boomsma 2009), most species appear to engage in multiple mating, either with one or both sexes remating. Selective breeding strategies frequently take advantage of the establishment of full- and half-sibling families, wherein polyandry or, more commonly, polygyny is imperative. In polyandry, females mate multiple times during their lifetime, or their eggs are fertilised by multiple males, resulting in multiple paternity within offspring from a single oviposition event. These reproductive strategies impact our capacity to infer genetic and environmental influences on traits.

To ensure genetic monogamy, the mating process must be meticulously controlled to prevent a female from mating more than one male. To establish paternal half-sibling families, males must have the ability to mate with several females. In black soldier fly, females are capable of remating after oviposition (Samayoa et al. 2016) and Hoffmann et al. (2021) uncovered evidence of multiple paternity, i.e., polyandry, meaning a mated female cannot simply be picked from a mass-reared population and assumed to produce full-sibling offspring. For the house fly, most females mate only once (Riemann et al. 1967), while males can mate multiple times (ter Haar et al. 2023). During mating, females receive a dose of seminal substance that stimulates oviposition and induces lifelong female refractoriness (Riemann et al. 1967). However, polyandry is possible if a mating is incomplete (female does not receive the full dose of seminal substance during copulation) (Arnqvist & Andrés 2006). These examples underscore the importance of gaining a thorough understanding of a species' basic reproductive strategy before embarking on genetic optimisation, to avoid significant waste of time and resources.

### *Mating behaviour*

Mating behaviour is important when artificial insemination is not an option. For instance, black soldier flies engage in lekking behaviour, wherein males aggregate at specific sites called leks (Tomberlin & Sheppard 2001), which is difficult to accommodate in small rearing environments. Even when the mating process can be controlled, there may be adverse consequences to consider. Firstly, reduced sexual selection could alter reproductive investments, potentially impeding population fitness and the success of breeding programs (e.g., Sorci et al. 2021). Secondly, interference with the timing of mating could reduce fecundity; for example, delaying mating might result in resources being diverted from egg production to longevity (Tomberlin et al. 2002). Thirdly, manipulating the environmental conditions in which mating occurs, including adjusting the male-to-female ratio and cage density, could influence mating behaviour and consequently affect mating success (Carrillo et al. 2012, Nakamura et al. 2016). A reduction in reproductive success should therefore be anticipated in insect breeding programs when wild or mass bred populations are transferred to controlled rearing environments.

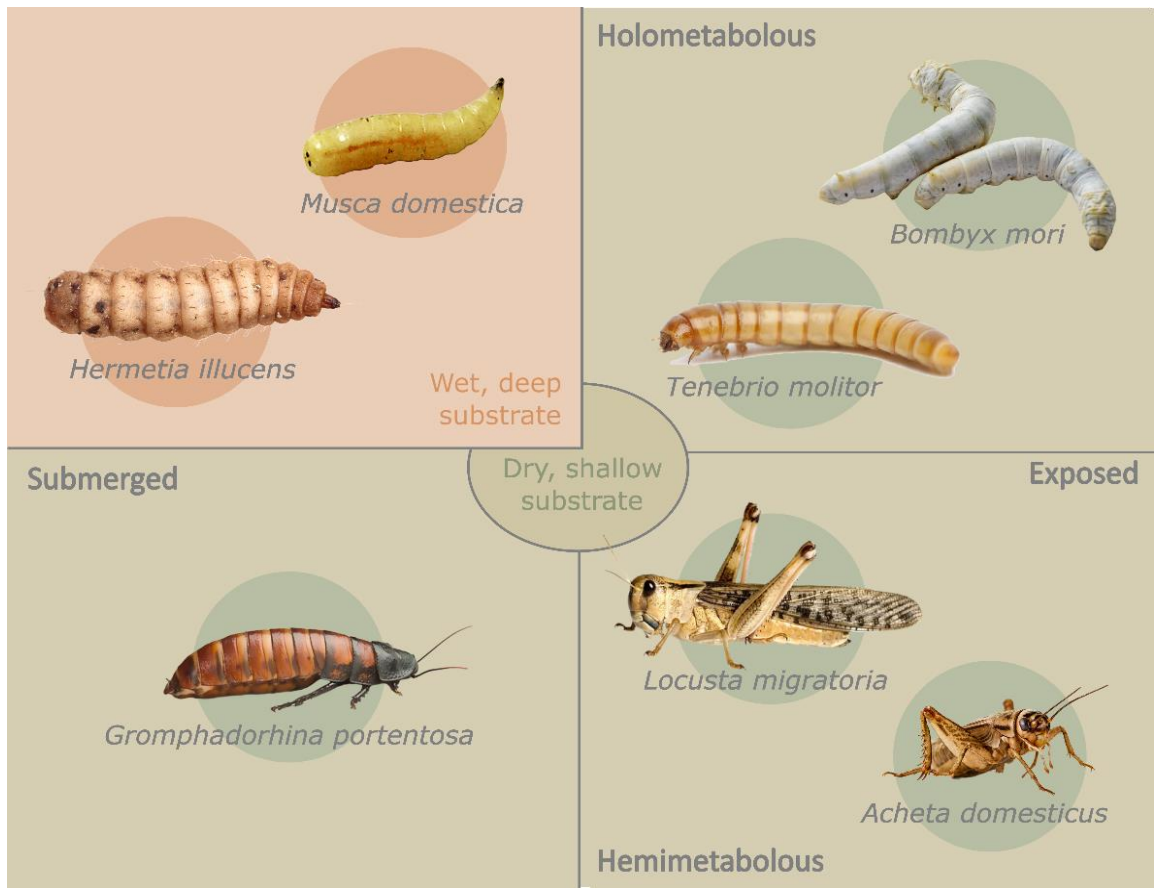

**Figure 3** Examples of insects reared globally for human food and/or animal feed and their biological properties: holometabolous or hemimetabolous life cycle, preference to being exposed or submerged in the rearing substrate or visible during rearing, and the properties of the standard rearing substrate. *Hermetia illucens*, *Musca domestica*, *Bombyx mori*, *Tenebrio molitor* and *Acheta domesticus* are authorised in the EU for feed purposes. *Locusta migratoria*, *Tenebrio molitor* and *Acheta domesticus* are authorised as novel foods. *Gromphadorhina portentosa* is not authorised in the EU for either purpose.

### *Tracking individuals in insect populations*

A crucial factor contributing to the success of selective breeding in traditional livestock species is the ability to trace individual performance across population pedigrees. Various techniques are employed to differentiate between individuals, including the use of ear tags, tattoos, clippings, and genomic information. Methods for distinguishing individual insects have also been developed (Walker & Wineriter 1981, Hagler & Jackson 2001) for example application of acrylic paint (as demonstrated by Jones and Tomberlin 2020) and tags affixed to the thorax using nail polish (Samayoa et al. 2016). In house flies, fluorescent dust colours have been used for marking (Meffert and Bryant 1991) and physical tags are commonly employed in honeybee breeding (Smith et al. 2021). Methods based on computer vision used to classify insect species (e.g. Bjerge et al. 2023, Yasmin et al. 2023) could potentially be adapted to within-species individual recognition or used to monitor and track individuals in a population (Nawoya et al. 2024). Despite these efforts, physical marking methods are often labour-intensive, time-consuming, and cannot be consistently applied and transferred across life stages due to shedding of the exoskeleton during growth. Holometabolous species pose unique challenges due to the substantial morphological distinctions between their larval, pupal, and adult stages, complicating the use of image-based identification. These aspects render current tagging and tracking options inadequate for retaining pedigree information and linking traits expressed in an individual at different life stages.

One potential solution is to rear each individual in isolation (Samayoa et al. 2016, Cammack & Tomberlin 2017, Cheon et al. 2022). When dealing with a large number of individuals, such procedures demand significant time for manual handling, extensive space, and specialised equipment for downscaled rearing, such as appropriate rearing containers which differ from those commonly used in production settings. Moreover, prolonged isolation can have adverse effects on various aspects of behaviour and performance (McCarthy et al. 2015, Vora et al. 2022). Changes in rearing density are known to influence juvenile development time and body weight, as observed in the black soldier fly and house fly (Parra Paz et al. 2015, Barragan-Fonseca et al. 2018, Kökdener & Kiper 2021). Consequently, traits measured in isolation may not accurately reflect those observed when insects are reared in larger groups as in production settings, introducing genotype by environment interaction (GxE). Individual rearing can also lead to increased mortality, particularly during early life stages (Samayoa et al. 2016), which can be detrimental to a breeding population, increase inbreeding and reduce genetic variance. An alternative approach is to rear entire families, or smaller groups of family members, in separate containers, enabling the maintenance of pedigree information until adulthood.

## Collecting phenotypic information

Collecting data on a breeding population serves as the foundation of selection decisions and, consequently, the entire breeding program. Historically, before genotyping became readily accessible for breeding purposes, phenotypic data was the sole information source. The prolific reproductive capacity, short life cycle, and small size of many insects pose challenges for efficient collection of phenotypic records. These challenges include low capacity to phenotype a representative sample of a large population within a short time frame, high technical error rates, labour intensity, observer biases, and risk of phenotyping being invasive. Consequently, phenotyping protocols characterised by both high accuracy and throughput should be developed for the species and traits of interest. For certain traits such as body mass, nutritional composition, growth rate or fecundity, group-level measurements are an alternative to individual records (Table 1). Phenotyping at group level reduces the information on within-group variance but may be the only option if time does not allow individual phenotyping, or if phenotyping is invasive.

**Table 1.** Examples of phenotyping of body size and egg production traits, the method, and level of phenotyping for commercial production insects.

| Trait | Method | Level | Species | Reference |
|---|---|---|---|---|
| **Body mass** | Weighing of fresh or dried specimen of immature or mature life stages | Individual | Black soldier fly *Hermetia illucens* | Meneguz et al. 2018, Shumo et al. 2019 |
| | | | House cricket *Acheta domesticus* | Ryder & Siva-Jothy 2001, Booth & Kiddell 2007 |
| | | | House fly *Musca domestica* | Boatta et al. 2023 |
| | | | Yellow mealworm *Tenebrio molitor* | Morales-Ramos et al. 2022 |
| | | Group | Black soldier fly *Hermetia illucens* | Yang & Tomberlin 2020, Scieuzo et al. 2023 |
| | | | Yellow mealworm *Tenebrio molitor* | Adamaki-Sotiraki et al. 2023 |
| **Body surface area** | Image analysis software for size estimation of live larvae | Individual | Black soldier fly *Hermetia illucens* | Laursen et al. 2021 |
| | | | House fly *Musca domestica* | Laursen et al. 2021, Hansen et al. 2024 |
| **Egg mass** | Weighing | Individual | Black soldier fly *Hermetia illucens* | Bertinetti et al. 2019 |
| | | Group | Black soldier fly *Hermetia illucens* | Bertinetti et al. 2019 |
| | | | House fly *Musca domestica* | Pastor et al. 2011 |
| | | | Yellow mealworm *Tenebrio molitor* | Adamaki-Sotiraki et al. 2023 |
| **Egg count** | Counting | Individual | Black soldier fly *Hermetia illucens* | Chia et al. 2018 |
| | | | House cricket *Acheta domesticus* | Nava-Sánchez et al. 2014 |
| | | | House fly *Musca domestica* | Khan et al. 2012, Francuski et al. 2020 |
| | | | Yellow mealworm *Tenebrio molitor* | Drnevich et al. 2001, Rho & Lee 2016 |
| | | Group | Black soldier fly *Hermetia illucens* | Shumo et al. 2019 |
| | | | Yellow mealworm *Tenebrio molitor* | Morales-Ramos et al. 2019 |

Another challenge pertains to linking trait measurements from juvenile and adult life stages. This is necessary to estimate trait correlations at the individual level, or to record sexually dimorphic traits. In many insect species, sex determination is not possible until adulthood, but traits of interest are often recorded at the juvenile life stages. Given that females generally exhibit larger body size than males (Stillwell et al. 2010, Teder & Kaasik 2023), selecting directly for size might lead to a skewed sex ratio towards females. Such an imbalance can lead to altered reproductive success which can result in an improvement (Hoc et al. 2019) or deterioration (Carrillo et al. 2012) of productivity depending on species and extremity of the sex ratio.

There is substantial untapped potential for implementing advanced automation, computer vision, or other sensor-based methods for phenotyping (Nawoya et al. 2024). Such developments have the potential to significantly enhance the speed and accuracy of data acquisition, ideally in real-time and with reduced influence from observer bias and handling effects. Technology-based methods may even become sensitive enough to discern subtle differences among sexes during juvenile life stages (Tao et al. 2019), a task that is often unfeasible or impossible using manual data collection methods. While phenotyping technologies used for individual size estimation and counting methods have already been developed for some insect species (Mallard et al. 2013, Duckworth et al. 2019), advancements on non-model commercial insect species remains limited. Examples of progress in this area is the assessment of larval size and thermal tolerance in house fly and black soldier fly using image analysis software (Laursen et al. 2021), the estimation of size distribution in a yellow mealworm population using image segmentation analysis (Baur et al. 2022) and the sex determination and counting of house cricket using object detection and classification techniques (Hansen et al. 2022). The implementation of methods that rely on a higher degree of automation and permit the assessment of multiple traits on the same individuals would substantially enhance the throughput of phenotyping while also facilitating multi-trait selection.

## Estimation of genetic parameters and genetic evaluation

Separation of important sources of variance, and the genetic parameters derived from those variance components, are used to predict genetic responses to selection and predict the genetic value (estimated breeding value, EBV) of an individual. The value of an individual's own phenotype for estimating EBVs depends on the proportion of phenotypic variance explained by additive genetic variance. In many instances, incorporating the performance of relatives improves the accuracy of selection (Figure 2) and thus the effectiveness of the breeding program.

### Collecting phenotypes from related individuals

Estimating narrow-sense heritability ($h^2$), trait correlations and EBVs entails two primary challenges: acquiring information on related individuals and disentangling genetic and environmental influences on the traits of interest. Traditional approaches to address these challenges are well-described in classical quantitative genetics theory (Falconer & Mackay 1996). Several experimental designs have been described facilitating the collection of phenotypic data on individuals with various types of relationships, and examples of their use in commercial insect species are presented in Table 2. To quantitative geneticist these designs may seem outdated, but they are highly relevant for insect research due to the massive challenges related to tracking information on relatedness.

### Parent-offspring records

The parent-offspring design utilises the covariance between the phenotypes of parents and their offspring to estimate heritability. This design has been effectively employed for the estimation of genetic parameters in the house fly and yellow mealworm (Table 2). The parent-offspring design necessitates the collection of phenotypic data on one or both parents and their offspring and thus requires the ability to establish maternal and/or paternal offspring. This can be accomplished by isolating mating pairs and subsequently isolating their offspring. The correlation between the phenotypes of fathers and their offspring provides the least biased estimate of narrow-sense heritability (Falconer & Mackay 1996). However, it may be more practical to track relatedness between mothers and offspring. One potential limitation of this design is that parents and their offspring require nearly identical rearing environments to accurately estimate the correlation between parent and offspring genotypes. The parent-offspring regression further assumes no non-genetic maternal effects, paternal effects, or epistasis, which are assumptions that are likely to be violated.

### Sibling records

Depending on the species, obtaining phenotypic records within a single generation rather than across generations may be more feasible. In such cases, data can be collected on full-siblings (full-sib design), half-siblings (half-sib design), or both (full-sib/half-sib design). Given the fecundity of most insect species, a design exclusively based on half-sibling relationships would seldom be relevant, as females generally lay multiple eggs in a single oviposition event.

Leveraging the covariance between siblings due to shared ancestry allows for the estimation of heritability and genetic correlations between traits. Such designs have been applied in the black soldier fly, house fly, house cricket, yellow mealworm and silkworm (Table 2). The implementation of the full-sib design may be the most straightforward for many insect species, as it only requires the isolation of females for oviposition and subsequently tracking or isolating full-siblings (Gray & Cade 1999, Bégin & Roff 2002, Bouwman et al. 2022). However, it is important to note that the full-sib design bias estimated genetic parameters (Falconer & Mackay 1996) and does not account for dominance and maternal effects. The paternal full-sib/half-sib design allows for an unbiased estimate of the additive genetic variance, but it necessitates polygyny. It is crucial to avoid multiple paternity within a single egg clutch (group of eggs produced in a single oviposition event) to prevent misinterpreting variance between half-siblings as full-sibling variance, which may require tighter mating control in some species (Gray & Cade 1999). Obtaining sufficient data to estimate variance components is a challenge when using the full-sib/half-sib design, particularly due to the short life cycle of many insects. Strategies such as phenotyping in several batches (Messina 1993) or decreasing temperatures during development to expand the development time (Gray & Cade 1999) can make the process more manageable, but may introduce GxE effects, challenging interpretation.

**Table 2** Examples of traits in commercial insect species where genetic parameters (narrow-sense heritability ($h^2$), phenotypic ($r_p$) or additive genetic ($r_a$) correlations) have been estimated, and the type of experimental design used for data collection.

| Trait | Species | Parameter | Design | Study |
|---|---|---|---|---|
| Courtship traits | House fly *Musca domestica* | $h^2$ | Parent/offspring regression | Meffert 1995 |
| Adult morphometrics | House fly *Musca domestica* | $h^2$ | Parent/offspring regression | Bryant & Meffert 1998 |
| Immune function Body size | House cricket *Acheta domesticus* | $h^2$, $r_a$ | Full-sib/half-sib design | Ryder & Siva-Jothy 2001 |
| Adult morphometrics | House cricket *Acheta domesticus* | $h^2$, $r_a$ | Full-sib/half-sib design | del Castillo 2005 |
| Body mass Spermatophore retention time | House cricket *Acheta domesticus* | $h^2$, $r_p$ | Full-sib/half-sib design | Mautz & Sakaluk 2008 |
| Immune function Elytra length (size) Cuticle melanism Development time | Yellow mealworm *Tenebrio molitor* | $h^2$, $r_p$, $r_a$ | Full-sib/half-sib design | Prokkola et al. 2013 |
| Cocoon length Cocoon weight Shell weight | Silkworm[*] *Bombyx mori* | $h^2$, $r_p$, $r_a$ | Full-sib design | Zambrano-Gonzalez et al. 2022 |
| Larval body mass Larval dry weight (pre)pupal body mass Development time | Black soldier fly *Hermetia illucens* | $h^2$, $r_p$ | Full-sib design | Bouwman et al. 2022 |
| Larval body weight Development time | Yellow mealworm *Tenebrio molitor* | $h^2$ | Parent/offspring regression | Morales-Ramos et al. 2022 |
| Larval dry weight Larval fat content | House fly *Musca domestica* | $h^2$ | Full-sib/half-sib design | Boatta et al. 2023 |
| Larval size Development time Larval survival Adult survival | House fly *Musca domestica* | $h^2$, $r_p$, $r_a$ | Full-sib/half-sib design | Hansen et al. 2024 |
| Larval weight Development time Larval protein content | Black soldier fly *Hermetia illucens* | $h^2$ | Full-sib design | Bouwman et al. 2024 |
| Reproduction Larval growth Larval survival Pupation rate Development time | Yellow mealworm *Tenebrio molitor* | $h^2$, $r_p$, $r_a$ | Full-sib design | Sellem et al. 2024 |

[*] See Hemmatabadi et al. 2016 for a review on silkworm genetic parameter estimation.

*Full-sib test groups*

The high fecundity observed in insects presents a distinct advantage, as it allows for the utilisation of full-sib phenotypic records in situations where recording all required information on single individuals is not feasible (Li & Margolies 1993). This becomes particularly relevant in cases involving body composition traits such as protein or lipid content, which necessitate culling individuals. Challenge testing on sib groups is common in aquaculture for carcass traits (Gjedrem 2010) and disease resistance traits (Ødegård et al. 2011), which are measured on the siblings of selection candidates. Phenotypic records collected from full siblings can also be leveraged to estimate genetic correlations between traits recorded at different life stages. In this context, it is crucial to account for other sources of covariance, such as maternal or environmental effects.

*Group records*

As described previously, phenotypes can be collected on groups. For selection, utilising full-sib averages or sums require a "pseudo" pedigree since full siblings share their parents. The loss of information occurs only within full-sib families where within-family variance is either unknown or unutilised. Maintaining a pseudo pedigree using full-sib family records further allows pairing males with multiple full-sib females for mating, since the females share both their phenotypic records and their pedigree. Providing a male with multiple females simultaneously can be stimulating for mating behaviour in some insect species (Tomberlin & Sheppard 2001). Several studies have explored the use of group records for variance component estimation in various species, including chicken, fish, mink, and rabbit (Simianer & Gjerde 1991, Gjøen et al. 1997, Biscarini et al. 2008, Biscarini et al. 2010, Nurgiartiningsih et al. 2010, Cooper et al. 2010, Peeters et al. 2013, Shirali et al. 2015, Piles and Sánchez 2019). Genetic models have also been developed for this purpose (Olson et al. 2006, Su et al. 2018, Gao et al. 2019). For insects, the group composition typically cannot be more diverse than consisting of full siblings while still preserving information on relatedness, unless individuals or families can be tracked in the rearing environment. This limitation introduces potential biases due to shared environmental effects, as discussed below.

*Common environment and maternal effects in insects*

The covariances observed both between individuals and between records of different traits measured on the same individual are influenced by a combination of genetic and non-genetic factors. Insects, being ectotherms, are highly susceptible to environmental influences (especially temperature), and the covariance of phenotypes among grouped individuals can be inflated due to shared environments, even when reared under standardised laboratory conditions (Hansen et al. 2024, Sellem et al. 2024). As groups of insects reared together in a selective breeding context would typically consist of full siblings, the separation of genetic and common environmental effects becomes especially important. To quantify and correct for common environmental effects it is necessary to replicate the sibling groups in multiple environments (Kristensen et al. 2005, Roff & Fairbarn 2011, Bouwman et al. 2022). Shared environmental effects are especially influential during the larval stage of holometabolous insects, where factors such as co-digestion, substrate consistency, and microenvironment play a crucial role (Gregg et al. 1990, Watson et al. 1993, Larraín & Salas 2008, Kökdener & Kiper 2021, Muurmann et al. 2024). Full siblings may share other sources of variance, such as litter effects or maternal non-genetic effects (Mousseau and Dingle 1991). In the vinegar fly (*Drosophila melanogaster*) transgenerational non-genetic maternal effects have been identified, including age effects (Hercus & Hoffmann 2000, Miller et al. 2014, Mossman et al. 2019, Lee et al. 2019) and dietary restriction effects (Lee et al. 2023). These effects can be included in the genetic model if recorded, but it is unfeasible to distinguish maternal effects from genetic effects shared by full siblings within a single egg clutch. Therefore, it is essential that breeding schemes for insects are designed and managed to minimise the influence of non-genetic effects, such as maternal age, nutritional environment, and density. A rigorous schedule for mating and egg collection, as well as meticulous monitoring of eclosion times to achieve age synchronisation among individuals, will minimise such effects.

### The genetic model

The mixed model approach is commonly used in animal breeding for variance component and EBV estimation, as it can simultaneously accommodate many types of genetic relationships and correct for systematic effects. This approach has also found application in insects (Roff & Fairbairn 2011, Sellem et al. 2024). Since the life cycle of most insect species follows a discrete pattern where each generation comprises individuals with thousands of close relatives, the benefits of utilising various types of relationships are most pronounced in populations with a multigenerational pedigree. The mixed model approach further offers a solution to address highly imbalanced family sizes and uneven contributions from parents, which are common challenges encountered in insect breeding. It also allows modelling both genetic and non-genetic influences on the traits of interest, thereby revealing environmental factors that may impact breeding and management decisions. Variance components are typically estimated using restricted maximum likelihood, and prediction of EBVs is based on best linear unbiased prediction (Henderson 1950, 1975). While the estimation of EBVs in farmed insects is not common outside of honeybee breeding, examples can be found in silkworm (Shabdini et al. 2011) and in evolutionary genetics (Roff & Fairbarn 2011). Similar to variance component estimation, EBVs can be predicted from group-level information when maintaining individual-level information is impractical or impossible (Olson et al. 2006, Biscarini et al. 2008 and 2010, Su et al. 2018, Ma et al. 2021). When using group information, a loss of prediction accuracy is expected, leading to a reduction in genetic gain. Combining group and individual information for EBV estimation could improve accuracy, which is possible if mating occurs shortly after adult phenotyping, allowing adults to be reared individually for a limited period. Including an individually recorded trait that is genetically correlated with the group-recorded traits has been shown to improve EBV estimation from group records (Ma et al. 2021).

## Selective breeding strategies

### *Phenotypic selection*

Phenotypic selection, often referred to as mass or individual selection, is a selection strategy where individuals are evaluated based on their own phenotypic performance. This approach is successfully used in aquaculture breeding for traits like growth rate (Gjedrem & Rye 2018) and has also found application in black soldier fly (Facchini et al. 2022, Hull et al. 2023), silkworm (Seidavi et al. 2014) and yellow mealworm breeding (Morales-Ramos et al. 2019, Song et al. 2022). The most significant advantage of phenotypic selection is its simplicity, as it eliminates the need for tracking pedigrees or individuals, which is a substantial challenge in insect breeding. All breeding animals can be reared together without the need for isolation or downscaling the environment. The important disadvantages of phenotypic selection can be found in Table 3. Phenotypic selection is an attractive option in insect productions in situations where selection on a single trait is the goal, the trait is directly observable on an individual, and has high heritability.

***Pedigree based selection***

When information on relatedness is available, several selection strategies become possible. Common to those is that they utilise pedigree information to predict EBVs which can increase selection accuracy and enable both single and multi-trait breeding, accounting for underlying genetic dependencies between traits in the breeding goal. Pedigree-based selection further allows multistage selection where different information sources can be utilised at different time points.

*Between-family selection* utilises the average performance of individuals within a family to estimate the family EBV, and families are subsequently ranked accordingly (Lush 1937). Entire families may either be selected or culled, with the flexibility to select all individuals within a family or only a specified number. This strategy only requires tracking the pedigree at family level, which is feasible for those insect species which can be reared in full-sibling groups. *Within-family selection*, however, requires individual tracking within a group, unless selection happens directly after phenotyping which does not accommodate EBV estimation. Combining between- and within-family selection has useful application in cases where environmental effects strongly influence both individuals and families. Another strategy which uses pedigree information is *Progeny selection.* Since the EBV of an individual reflects the performance of their offspring, using the actual progeny performance as input gives very accurate estimates of EBVs. The strategy is only applicable in insect species that do not experience discrete generations, and where selection can be postponed until offspring have been evaluated. As described previously, females of some insect species utilised for food and feed mate only once, wherefore selecting breeding stock based on progeny performance is equivalent to selecting the offspring directly. For most commercially produced insect species, this strategy is currently not applicable.

**Table 3** Examples of different selection strategies, the information sources they use, the advantages and disadvantages of their implementation and their applicability in insect breeding plans.

| Selection strategy | Information source(s) | Advantages | Disadvantages | Insect breeding |
|---|---|---|---|---|
| *Phenotypic (mass) selection*<br><br>Selection based on own phenotype | Own phenotype | • Simple to implement<br>• No individual or pedigree tracking required<br>• Efficient for traits with moderate to high heritability e.g. size | • Inefficient for low-heritability traits<br>• No inbreeding control<br>• Multi-trait selection inefficient<br>• Traits should be measurable on selection candidates | • Diminishes the need to track individuals and has been proven efficient for single-trait selection in insects |
| *Family selection*<br><br>Selection based on family performance | Phenotypes<br><br>Pedigree | • Balances selection intensity and genetic diversity<br>• Sib-groups can be used for challenge tests and invasive phenotyping<br>• Useful for low heritability traits in species with high reproductive rate | • Limited genetic variation within families<br>• Disregard within-family variation<br>• Large number of families required<br>• Risk high inbreeding rate<br>• Confounding by common environment | • Tracking families instead of individuals is feasible in insects<br>• Common environment effects can be substantial |
| *Progeny selection*<br><br>Selection based on progeny performance | Phenotypes<br><br>Pedigree | • Suitable for low heritability traits, traits measured in one sex, or slaughter traits<br>• Gives accurate estimates of breeding values<br>• Rapid genetic gain, targeted improvement<br>• Allows multi-trait selection | • Requires pedigree and performance data<br>• Increases generation interval<br>• Costly | • Tracking progeny is a challenge<br>• Does not allow discrete generations<br>• Does not allow short life span of adults<br>• Equivalent to selecting offspring directly if parents cannot remate |
| *Crossbreeding*<br><br>Selection in purebred lines to improve crossbred offspring | Phenotypes<br><br>Pedigree | • Utilises heterosis<br>• Efficient when traits in breeding goal are unfavourably correlated<br>• Possible to cross purebred lines bred for sex-specific traits | • Risk mating incompatibility between purebred lines<br>• Heterosis effects diminish over time<br>• Risk low trait uniformity within a crossbred line and outcomes are unpredictable<br>• Time consuming | • Housing requirements may be substantial when maintaining purebred and crossbred lines. |
| *Genomic selection*<br><br>Selection based on genetic merit derived from genotype information | Reference population<br><br>Marker genotypes | • Increases selection accuracy and speed<br>• Reduces generation interval<br>• Effective for low heritability traits | • Requires single nucleotide polymorphism (SNP) array<br>• High costs for genomic data<br>• Time is required for generating and analysing genotype data | • Cost of genotyping individual insects is high<br>• Time required may be a limiting factor<br>• Generation interval is short in many insects of interest |

*Crossbreeding*

Crossbreeding, where purebred parents from two genetically distinct lines are mated, can be efficient in producing superior offspring due to heterosis. Heterosis arises from an increased heterozygosity in offspring compared to parental lines. The degree of heterosis depends on differences in allele frequencies in loci contributing to trait variation between individuals (McAllister 2002), although large differences simultaneously increase the risk of outbreeding depression. When crossbreeding is used solely for crossing two different lines to utilise heterosis, additive genetic effects do not change. Heterosis effects are reduced in later generations, wherefore it should be combined with selection within a purebred line itself (e.g. Bowman 1959). Smith (1964) suggested to use sire and dam lines in which different traits were selected for. This has been implemented both in pig and poultry breeding. Due to traits which are typically negatively genetically correlated such as growth rate and development time (Roff 2000), selecting for specialised lines could also potentially be used here.

Crossbreeding has not been applied systematically in insects outside of silkworm production (Strunnikov 1986). Adamaki-Sotiraki et al. (2023) crossed four strains of yellow mealworm originating from different geographical areas to study mating compatibility. All crosses had compatible matings but there were no differences in number of eggs or larval survival between crosses and pure lines, meaning no heterosis was documented. Mating compatibility should always be tested before starting a crossbreeding program to avoid adverse effects on population production.

## Utilising genomic information in insect breeding

Obtaining genomic information within a breeding program offers several significant advantages. Primarily, as emphasised throughout this review, the precise identification and tracking of individuals and their familial relationships pose considerable challenges in implementing any insect breeding program. Genetic markers provide a highly accurate mean to address this issue. In a manner analogous to many traditional agricultural species, the assignment of individuals to specific families or lines can be accomplished through single nucleotide polymorphism (SNP) array genotyping (Calus et al. 2011, Hayes 2011, Huisman 2017). A SNP array, constructed from polymorphic variants identified through whole-genome sequencing, can effectively cluster individuals based on their degree of relatedness. This gives the opportunity for communal rearing of families avoiding bias from the common environmental effects. Furthermore, the SNP array can serve as a valuable tool for monitoring genomic inbreeding within commercial insect populations. While traditional population pedigrees offer estimates of population-level inbreeding, genomic data allows for the monitoring of realised inbreeding at the individual and population levels, often quantified through the assessment of runs of homozygosity (ROHs). Additionally, genomic data provides insights into the level of genomic variation within a population, enabling informed population management and breeding decisions (Hoffmann et al. 2021, Generalovic et al. 2023). Several prerequisites must be met for genotyping to be both feasible and beneficial within an insect breeding context. First, high-quality genome assemblies are imperative for the species utilised in production (Eriksson et al. 2020, Zhan et al. 2020, Generalovic et al. 2021, Dossey et al. 2023). Additionally, to develop a universally applicable SNP array across various populations, a high number of high-density genome-wide molecular markers that demonstrate polymorphism across multiple populations is essential. Utilising genomic resources from existing population genetic studies can be instrumental in designing effective SNP chips (e.g., Bahrndorff et al. 2020, Kaya et al. 2021).

Another application of genotype information is for the implementation of genomic selection (Meuwissen et al. 2001), widely adopted in the selection of breeding candidates across many agricultural species (Goddard et al. 2010, Crossa et al. 2014, Rutkoski et al. 2014). Genomic selection has yielded higher rates of genetic gain compared to pedigree-based selection methods. For instance, in dairy cattle, genomic selection has doubled the rate of genetic gain (García-Ruiz et al. 2016, Doublet et al. 2019). This increase in genetic gain primarily results from the ability to predict EBVs with greater accuracy at a very early stage in an individual's life, leading to a substantial reduction in the generation interval followed by an increase in selection intensity, since more selection candidates are available at a young age. The application of genomic selection in insect breeding does however not provide any obvious current advantages. A reduction of the generation interval could be beneficial in species with long life cycles, but for most commercial species the life cycle is short (less than 3 months) and this already creates practical challenges. Sexual maturity in insects is not reached until the adult life stage, so final selection decisions with restriction on number of selected males and females cannot be made until this stage, even if candidates can be preselected at juvenile life stages. The high fecundity of insects already enables high selection intensity, so an additional small increase in selection accuracy is only beneficial if it results in a reranking of candidate individuals or families. The time required for sampling, DNA extraction, library preparation, and bioinformatics analysis may exceed the available timeframe for obtaining and applying genomic selection. Additionally, the quantity of DNA obtainable from a single individual may prove insufficient for genotyping, contingent on the insect's life stage and species, potentially necessitating the sacrifice of the individual, thereby diminishing the overall benefits of acquiring genomic information. Alternative sampling methods, such as haemolymph, exoskeleton, wing, bristle, or other tissue samples, should be explored and their suitability for breeding program applications thoroughly assessed. Economic considerations may also render individual insect genotyping prohibitive, prompting alternative strategies like pooled genotyping (Ashraf et al. 2016, Ørsted et al. 2019). If genomic information can be successfully integrated into selective breeding schemes for commercial insects, it may offer solutions to some of the fundamental challenges associated with maintaining pedigreed populations and managing the adverse effects of inbreeding which currently spark concern in the field.

## Sustainable harnessing of genetic variation in insect production

A downside to the promising potentials of insect selective breeding is that genetic drift and rates of inbreeding progresses within closed populations and across generations. The rate of inbreeding increases inversely proportionally with the effective population size (Ne). Thus, the lower the Ne, the faster the loss of genetic variation and the higher the rate of inbreeding (Crow & Kimura 1970, Ohta & Kimura 1973). A high selection intensity will typically lower Ne as seen in many populations of traditional livestock species where effective population sizes are often in the range of 50 to a few hundred individuals (Leroy et al. 2013). Hence, these populations may experience rapid genetic improvement at the expense of increased genetic drift, resulting in a loss of genetic variation. Additionally, they risk reduced fitness due to the manifestation of inbreeding depression, which is exacerbated by the increased rate of inbreeding (Willi et al. 2022). Given the short generation intervals for most insects, these processes can develop rapidly (Hoffmann et al. 2021). If not managed, genetic drift and inbreeding can constrain selection responses, compromise fitness, and cause selection responses diverging from those intended in the breeding program.

### *Consequences of inbreeding and genetic drift in insect populations*

The background and extent of genetic drift and inbreeding are well described theoretically and in multiple model species in the laboratory, in traditional livestock, as well as in wild populations of conservation concern (Kristensen & Sørensen 2005, Willi et al. 2022). However, little is known about the level and impact of inbreeding in commercially reared insect populations to be used as food and feed (Rhode et al. 2020). Typically, the origin of the populations used in studies on e.g. black soldier fly or yellow mealworm are not described in detail in published literature. First reports on selective breeding employ phenotypic selection (Facchini et al. 2022), which can lead to very fast and marked increases in inbreeding (Bentsen & Olesen 2002, Hill 2014, Hoffmann et al. 2021, Fan et al. 2022). Considering the rapid and significant genetic deterioration that can result from high levels of population inbreeding, it is imperative that this factor is given thorough and serious consideration within the context of this emerging industry.

Reduced fitness due to inbreeding is well-documented (Frankham et al. 2002, Kristensen et al. 2015). The degree of inbreeding depression for a particular trait differs among individuals and populations and is typically more pronounced for traits closely associated with fitness (Wright et al. 2008). Hoffmann et al. (2021) described loss of genetic variation and high rates of inbreeding in commercial black soldier fly populations, and although causation is typically lacking, anecdotal evidence of population collapses of commercial insect populations may be linked to loss of genetic variation and/or inbreeding. However, not all inbred populations suffer from inbreeding (Robinson et al. 2018). Such examples have led some authors to speculate that genetic problems in natural populations might be minor relative to demographic and environmental risks (Teixeira & Huber 2021). However, there is overwhelming empirical evidence that inbreeding depression is often substantial in natural and domestic populations, and the few exceptions of populations that thrive despite being highly inbred is not in conflict with well-established theory and should not dictate recommendations. Reasons why some highly inbred individuals and populations seem unaffected by inbreeding include stochastically low genetic load segregating in populations and purging of deleterious mutations (Kimura et al. 1963, Willi et al. 2022).

An added level of complexity arises from the environmental dependence of both inbreeding depression and purging. Favourable environmental conditions can moderate the expression of inbreeding depression, with the flip side that inbreeding effects are typically more severe in harsh environments (Reed et al. 2012). Thus, insect breeders might experience sudden inbreeding related collapses in periods with suboptimal diets or following disease outbreaks. These considerations hold particular significance in the context of commercial insect production for food and feed as one argument supporting the environmental sustainability of this production method is the capacity to rear insets on waste products with low nutritional value (van Huis & Tomberlin 2017). Additionally, high temperature stress can constitute a challenge in larvae with fast growth (Li et al. 2023) in some commercial management systems, exemplifying the importance of inbreeding by environment interactions in insect production systems.

Establishing diverse commercial base populations, maintaining high Ne and considering crossbreeding and potential continuous inflow of variation from wild populations are key to a long-term sustainable breeding program. Selective breeding on pedigreed populations allows for monitoring and controlling inbreeding and genetic drift and although challenging, efforts are needed to overcome challenges related to establishing pedigreed insect populations. The biology of most insect species with large reproductive output per female, little space requirements, and limited capacity of individual males to mate many females (which contrasts to the situation in e.g. cattle where one ejaculate from a bull can be used to inseminate up to 400-500 females) all speaks in favour of the ability to administer inbreeding and genetic drift in a sustainable way.

### *Unfavourable responses to selection*
Phenotypic traits and their underlying genetic architectures are not independent. Pleiotropic loci or linkage leads to genetic correlations, where selection for one trait will lead to possibly unfavourably correlated responses in other traits or be counteracted by natural selection on correlated life history traits. An example of the latter is provided by the general trade-off between reproduction and immune defence in insects (Schwenke et al. 2016). Both processes are energy demanding, and artificially applied selection for reproductive output might be counteracted by natural selection on the immune response applied by pathogens. Thus, in addition to paying attention to managing inbreeding and maintaining genetic variation in domesticated insects, it is also important to consider the risk of inadvertent responses to selection, which can occur because of environmental and artificial selection pressures. Commonly, domestication leads to inadvertent selection for fast development and early reproduction, unless deliberately avoiding this. These traits are likely to be accompanied by small body size, and lower reproductive success (see discussion in Leftwich et al. 2016). Sgrò and Hoffmann (2004) found that genetic correlations in experimental evolutionary studies were not constant across different environments. An example was provided by Hillesheim and Stearns (1991) who found a correlated response in developmental time to selection for body mass in the vinegar fly, but only when selection was applied under poor nutrition.

It is difficult to precisely predict the outcome of correlated responses and inadvertent selection. However, it is worth keeping in mind that particularly energy demanding processes might come with a life history cost. On a positive note, not all life history costs are problematic in the production setting, i.e. we might readily accept decreased developmental time in our breeding stock, or decreased reproduction in the individuals destined for harvesting before the adult stage. Further, we also point to the review by Sgrò and Hoffmann (2004), which suggest that some genetic correlations might be broken under certain environmental conditions. Thus, we advise to consider possible correlations and inadvertently applied selection pressures when deciding on a selection strategy and find that this issue hold both challenges and opportunities for the future.

## Future considerations

The meticulous handling and manipulation of the rearing environment needed for successful implementation of selective breeding means that environments in the breeding and production units will differ. GxE interactions can thus impede the outcomes of selection if superior genotypes in breeding units are not performing well at the production level (Sandrock et al. 2022). This can be caused by an environment and population specific genetic basis of traits being selected for and evaluated (e.g. Mulder & Bijma 2005). Consequently, information obtained in genetic evaluations is not directly transferable between populations or across environments. This should be considered when evaluating the outcomes of a breeding scheme and is especially relevant for insects, as diets and other environmental conditions can vary markedly not only between breeding and production units, but also across farms and seasons.

As selective breeding becomes more widely applied in commercial insects, focus should also be on evolutionary trade-offs that may exist between e.g. production relevant traits and traits related to behaviour, stress tolerance and disease resistance (Fischer et al. 2005, Guerra & Pollack 2009, Zerjal et al. 2021). Such trade-offs are well known from traditional livestock animals as exemplified by behavioural and skeletal deformities in fast growing broilers (Julian 1998) and reduced fertility in high yielding dairy cattle breeds (Berry et al. 2003) (for reviews on the topic see Rauw et al. 1998 and van Marle-Köster & Visser 2021). Developing sustainable breeding goals and incorporating genetic correlations between production and behavioural/robustness traits into breeding programs is therefore key to producing populations equipped to withstand stress or diseases (Kovačić et al. 2020). Breeding for robustness could serve as a strategic approach to ensure long-term persistence of a population, mitigate the risk of population collapse, and ensure ethical breeding practices and improved welfare in the populations under selection (Rauw & Gomez-Raya 2015).

The use of insects in our agricultural systems remain constrained by the biological limits of the species we are currently farming, including their abilities as degraders, rearing requirements, and nutritional profiles. More than one million insect species are described and most of them have a wide natural distribution. However commercial insect production is currently focused on a handful of species and populations. Harnessing the natural diversity within and between species, combined with selective breeding practices, has the potential to be a game-changer in sustainable production of insect-based food and feed.

Quantitative genetic theory has been instrumental in optimising traits of interest in traditional livestock for decades. Animal breeding practices tailored to insects and used wisely will provide a strong and untapped foundation for increasing the effectiveness of sustainable insect-based protein production.

## Data availability

Not applicable

## Acknowledgements

This review was conducted as part of the project "Optimisation of insect production for animal feed through breeding" funded by the Independent Research Fund Denmark. We acknowledge the funding support in addition to all project participants and collaborators, and their contributions to discussions throughout the project, which ultimately framed the topics included in this review.

## Author contributions

All authors conceptualised the scope and contributed to the initial draft. LSH and HMN revised the draft. All authors reviewed and approved the final manuscript.

## Funding

This work was supported by the Independent Research Fund Denmark (DFF-0136-00171B and DFF-1127-00081B) and the Novo Nordic Foundation (NNF21OC0070910).